\documentclass{article}%
\usepackage{amsfonts}
\usepackage{amsmath}
\usepackage{amssymb}
\usepackage{graphicx}%
\begin{document}

\title{On the embedding of spacetime in higher-dimensional spaces with torsion}
\author{$^{a}$C. Romero, J. B. Formiga, L. F. P. da Silva, $^{b}$F. Dahia,\\$^{a}$Departamento de F\'{\i}sica, Universidade Federal da Para\'{\i}ba, \\C. Postal 5008, 58051-970 Jo\~{a}o Pessoa, Pb, Brazil\\$^{b}$Departamento de F\'{\i}sica,Universidade Federal de Campina\\\ Grande, 58109-970, Campina Grande, Pb, Brazil \\\ \ e-mail: cromero@fisica.ufpb.br, Brazil}
\maketitle

\begin{abstract}
We revisit the Riemann-Cartan geometry in the context of recent
higher-dimensional theories of spacetime. After introducing the concept of
torsion in a modern geometrical language we present some results that
represent extensions of Riemannian theorems. We consider the theory of \ local
embeddings and submanifolds in the context of Riemann-Cartan geometries and
show how a Riemannian spacetime may be locally and isometrically embedded in a
\textit{bulk} with torsion. As an application of this result, we discuss the
problem of classical confinement and the stability of motion of particles and
photons in the neighbourhood of branes for the case when the bulk has torsion.
We illustrate our ideas considering the particular case when the embedding
space has the geometry of a warped product space . We show how the confinement
and stability properties of geodesics near the brane may be affected by the
torsion of the embedding manifold. In this way we construct a classical
analogue of quantum confinement inspired in theoretical-field models by
replacing a scalar field with a torsion field.

\end{abstract}

\section{Introduction}

The idea that our spacetime may have more than four dimensions seems to be a
recurrent theme in contemporary theoretical physics research. Such idea was
first conjectured by G. Nordstr\"{o}m \cite{Nordstrom}, in 1914 (before the
completion of general relativity), whose aim was to achieve unification of
gravity with electromagnetism. Although Nordstr\"{o}m's interesting work, done
in the context of a scalar gravity theory, was ignored for a long time, the
same basic idea was taken up again, a few years later, by the mathematician T.
Kaluza \cite{kaluza}. Kaluza, assuming the existence of a fifth dimension, was
able to show that the basic equations of the gravitational and
electromagnetism fields could be derived from the Einstein equations written
in a five-dimensional space. Kaluza's theory, later known as\ the Kaluza-Klein
theory (after the contribution from physicist O. Klein) became the point of
departure of many higher-dimensional theories. In the seventies, the
Kaluza-Klein theory was generalized to include more general gauge fields
\cite{kaluza1}, which required the introduction of additional extra
dimensions. These\ new developments led to string theory in the eighties
\cite{superstrings}, while the nineties saw other higher-dimensional proposals
come to light. Among these we should mention D-brane theory, the \textit{brane
world scenario }\cite{Randall} and the so-called induced matter approach
\cite{Wesson}.

In almost all theories mentioned above it has been generally assumed that the
underlying higher-dimensional space (often referred to as \textit{the bulk
}\cite{Roy}) has a Riemannian character. Surely this is the more natural
assumption to be made since the Riemannian theory is the\ geometrical setting
of the well-established theory of general relativity. With very few
exceptions, there has not been much discussion\ on whether the bulk could
admit more general geometries. Nevertheless, attempts to broaden this scenario
started to appear recently in the literature. Non-Riemannian geometries, such
as Weyl geometry or Riemann-Cartan geometry, are taken into consideration as
viable possibilities to describe the bulk \cite{Israelit,Arias,Nandinii}.

The development\ of differential geometry in the last century led to the
discovery of a vast number of non-Riemannian geometries. The richness these
geometries possess in the form of new geometrical structures (in addition to
the metric and affine connection) render them rather apt to the formulation of
new physical theories insofar as they introduce extra degrees of freedom
suitable, for instance, for the description of non-gravitational physical
fields. Of course this is\ a well-known fact and was for a long time explored
by A. Einstein and others in their pursuit of a unified field theory
\cite{Gonner}.\ Nevertheless, we believe it is still of interest to
investigate non-Riemannian geometries in a more modern context, namely, that
of higher-dimensional\ spacetime theories. An illustrative example of
development in this direction has appeared recently, in which it is assumed
that, in the context of five-dimensional spacetime theory, the bulk has a
Weylian geometry. One of the results of such an approach is that one is able
to establish a classical analogue of quantum confinement \cite{Rubakov} by
purely geometrical means \cite{Gomez}.

Following the ideas mentioned above, we consider, in the present article,
another kind on non-Riemannian geometry, namely the Riemann-Cartan geometry.
The latter represents one of the simplest generalizations of Riemannian
geometry. As is well known, it constitutes the geometrical framework of a
theory formulated by E. Cartan \cite{Cartan} in an attempt to extend general
relativity when matter with spin is present. In spite of the limited interest
it has arisen among theoretical physicists since its conception (perhaps due
to the fact that it differs very little from general relativity), some authors
believe that the Einstein-Cartan theory can have an important role in a future
quantum theory of gravitation \cite{Trautman}. Moreover, torsion cosmology has
been investigated recently in connection with the acceleration of the Universe
\cite{Shie}.

The paper is organized as follows. We start in Section 2 with a brief review
of Riemann-Cartan geometry and then prove some mathematical results that
represent straightforward extensions of Riemannian theorems. We proceed in
Section 3 to consider the theory of \ local embeddings and submanifolds in the
context of geometries with torsion. Here we show that a Riemannian manifold
may be embedded in a higher-dimensional space with torsion and this
constitutes one of our main results. Section 4 contains an application of the
formalism to the problem of classical confinement and the stability of motion
of particles and photons in the neighbourhood of hypersurfaces. In Section 5
we show how the presence of a torsion field may affect both the confinement
and/or stablity of the particle's motion. In Section 6 we give a simple
application of the ideas developed previously. We conclude, in Section 7, with
our final remarks.\ \ \ \ \ \ \ \ \ \ \ \ \ \ \ \ \ \ \ \ \ \ \ \ \ \ \ \ \ \ \ \ \ \ \ \ \ \ \ \ \ \ \ \ \ \ \ \ \ \ \ \ \ \ \ \ \ \ \ \ \ \ \ \ \ \ \ \ \ \ \ \ \ \ \ \ \ \ \ \ \ \ \ \ \ \ \ \ \ \ \ \ \ \ \ \ \ \ \ \ \ \ \ \ \ \ \ \ \ \ \ \ \ \ \ \ \ \ \ \ \ \ \ \ \ \ \ \ \ \ \ \ \ \ \ \ \ \ \ \ \ \ \ \ \ \ \ \ \ \ \ \ \ \ \ \ \ \ \ \ \ \ \ \ \ \ \ \ \ \ \ \ \ \ \ \ \ \ \ \ \ \ \ \ \ \ \ \ \ \ \ \ \ \ \ \ \ \ \ \ \ \ \ \ \ \ \ \ \ \ \ \ \ \ \ \ \ \ \ \ \ 

\section{Riemann-Cartan geometry}

In this section we review some basic definitions and mathematical facts of
Riemannian and Riemann-Cartan geometry. As we shall see, the latter may be
viewed as a kind of generalization of the first, and some theorems that will
be presented here are straightforward extensions of corresponding theorems of
Riemannian geometry. However, these extensions present new features specially
as far as geodesic motion is concerned. Let us start with the definition of
affine connection \cite{do Carmo}.

\textbf{Definition}. Let $M$ be a $n$-dimensional\ differentiable manifold and
$T(M)$ the set of all differentiable vector fields on $M$. An \textit{affine
connection } is a mapping $\nabla:T(M)\times T(M)$\ $\rightarrow T(M)$, which
is denoted by $(U,V)\rightarrow\nabla_{U}V$, satisfying the following
properties:
\begin{equation}
i)\text{ }\nabla_{fV+gU}W=f\nabla_{V}W+g\nabla_{U}W, \label{affine1}%
\end{equation}%
\begin{equation}
ii)\text{ }\nabla_{V}(U+W)=\nabla_{V}U+\nabla_{V}W, \label{affine2}%
\end{equation}%
\begin{equation}
iii)\text{ }\nabla_{V}(fU)=V[f]U+f\nabla_{V}U, \label{affine3}%
\end{equation}
where $V,$ $U,$ $W\in T(M),$ and $f$, $g$ are \ $C^{\infty}$\ scalar functions
defined on $M.$ An important result comes immediately from the above
definition and allows one to define a covariant derivative along a
differentiable curve.

\textbf{Proposition}. \textbf{ }Let $M$ be a differentiable manifold endowed
with an affine connection $\nabla$, $V$ a vector field defined along a
differentiable curve $\alpha:(a,b)\subset R\rightarrow M$. Then, there exists
a unique rule which associates another vector field $\frac{DV}{d\lambda}$
along $\alpha$ to $V$, such that
\begin{equation}
i)\text{ }\frac{D(V+U)}{d\lambda}=\frac{DV}{d\lambda}+\frac{DU}{d\lambda},
\label{covariant1}%
\end{equation}%
\begin{equation}
ii)\text{ }\frac{D(fV)}{d\lambda}=\frac{df}{d\lambda}V+f\frac{DV}{d\lambda},
\label{covariant2}%
\end{equation}
where $\alpha=$ $\alpha(\lambda)$ and $\lambda$ $\in(a,b)$.

$\qquad\qquad iii)$ If the vector field $U(\lambda)$ is induced by a vector
field $\hat{U}$ $\in T(M)$, i.e., $U(\lambda)=\hat{U}(\alpha(\lambda))$, then
$\frac{DU}{d\lambda}=\nabla_{V}$ $\hat{U}$, where $V$ is the tangent vector to
the curve $\alpha$, i.e., $V=\frac{d}{d\lambda}.$ For a proof of this
proposition we refer the reader to \cite{do Carmo}. We now introduce the
concept of parallel transport of a vector\ along a given curve.

\textbf{Definition}. Let $M$ be a differentiable manifold \ with an affine
connection $\nabla$, $\alpha:(a,b)\in R\rightarrow M$ a differentiable curve
on $M$, and $V$ a vector field defined along $\alpha=$ $\alpha(\lambda)$.\ The
vector field $V$ is said to be \textit{parallel }if $\frac{DV}{d\lambda}=0$
for any value of the parameter $\lambda\in(a,b)$.

A concept that is basic to the Riemann-Cartan geometry is that of
\textit{torsion}, which is given by the following definition:

\textbf{Definition}. \textbf{ }Let $\nabla$\ be an affine connection defined
on $M$ \ and $U,V\in T(M)$. We define the \textit{torsion }$T$ of $M$ as the
mapping $T:T(M)\times T(M)$\ $\rightarrow T(M)$, such that
\begin{equation}
T(U,V)=\nabla_{U}V-\nabla_{V}U-[U,V]. \label{torsionless}%
\end{equation}
If the torsion vanishes identically we say that the affine connection $\nabla$
is \textit{symmetric} (or, simply, \textit{torsionless}).

To establish a link between the affine connection $\nabla$ and the metric $g$
we need a further definition. \qquad\qquad\ \ 

\textbf{Definition}. Let $M$ be a differentiable manifold \ endowed with an
affine connection $\nabla$ and a metric tensor $g$ globally defined in$\ M$.
We say that $\nabla$ is compatible with $g$ if for any vector fields $U,V,$
$W\in T(M)$,\ the condition below is satisfied:%

\begin{equation}
V[g(U,W)]=g(\nabla_{V}U,W)+g(U,\nabla_{V}W). \label{W-compatible}%
\end{equation}
We now state an important result.

\textbf{Theorem (Levi-Civita extended)}. In a given differentiable manifold
$M$ endowed with a metric $g$ on $M$, there exists only one affine connection
$\nabla$ such that $\nabla$ is compatible with $g$.

Proof. Let us first suppose that such $\nabla$ exists. Then, from
(\ref{W-compatible}) we have the following three equations%
\begin{equation}
V[g(U,W)]=g(\nabla_{V}U,W)+g(U,\nabla_{V}W)\label{1}%
\end{equation}%
\begin{equation}
W[g(V,U)]=g(\nabla_{W}V,U)+g(V,\nabla_{W}U)\label{2}%
\end{equation}
\begin{equation}
U[g(W,V)]=g(\nabla_{U}W,V)+g(W,\nabla_{U}V)\label{3}%
\end{equation}
Adding (\ref{1}) and (\ref{2}) and subtracting (\ref{3}), and also taking into
account the definition of torsion (\ref{torsionless}), we are left with
\begin{align}
g(\nabla_{V}W,U) &  =\frac{1}{2}%
\{V[g(W,U)]+W[g(V,U)]-U[g(W,V)]+g([V,W],U)\nonumber\\
&  +g([U,V],W)+g([U,W],V)-g(T(W,U),V)-g(T(V,U,),W)\nonumber\\
&  -g(T(W,V),U)\}\label{W}%
\end{align}
If the affine connection $\nabla$ is symmetric the above equation has a
simpler form, and in this case $\nabla$ reduces to the celebrated Levi-Civita
connection \cite{do Carmo}. The equation\ (\ref{W})\ shows that the affine
connection $\nabla$, if it exists, is uniquely determined from the metric $g$
and the torsion $T$ . ( In the torsionless case, $\nabla$ is determined from
$g$ alone). Finally, to prove the existence of such a connection we just
define $\nabla_{U}V$ by means of (\ref{W}).

A tensor that is naturally associated with $T$ is the \textit{torsion tensor}
$\mathcal{T}$ , defined by the mapping $\mathcal{T}$ \ $:T^{\ast}(M)\times
T(M)\times T(M)$\ $\rightarrow R$\ , such that $\mathcal{T}$ $(\widetilde
{w},U,V)=$\ $\widetilde{w}(T(U,V))$, where$\ T^{\ast}(M)$ denotes the set of
all differentiable one-form fields on $M$ and $\widetilde{w}\in\ T^{\ast}(M)$.
\ It is easy to see that the components of $\mathcal{T}$ \ in a coordinate
basis associated with a local coordinate system $\left\{  x^{a}\right\}  $,
$a=1,...,n$, are simply given in terms of the connection coefficients, i.e.,
$\mathcal{T}$ $_{\;bc}^{a}=\Gamma_{\;bc}^{a}-\Gamma_{\;cb}^{a}$, where
$\Gamma_{\;bc}^{a}\equiv dx^{a}(\nabla_{\partial_{b}}\partial_{c})$ . A
straightforward calculation shows that one can express the components of the
affine connection as%
\begin{equation}
\Gamma_{bc}^{a}=\{_{bc}^{a}\}-K_{\;bc}^{a}%
\end{equation}
where $\{_{bc}^{a}\}=$ $\frac{1}{2}g^{ad}[g_{db,c}+g_{dc,b}-g_{bc,d}]$ denotes
the Christoffel symbols of second kind and $K_{\;bc}^{a}=\frac{1}%
{2}(\mathcal{T}_{\;cb}^{a}+\mathcal{T}_{cb}^{\;\;a}+\mathcal{T}_{bc}%
^{\;\;a}),$ represents the components of another tensor, called the
\textit{contorsion tensor} \footnote{Note that the indices appearing in \ the
components of the torsion are raised and lowered with $g^{ab}$ and $g_{ab}$,
respectively.}.

Thus we see then that what basically makes the geometry discovered by Cartan
distinct from Riemannian geometry is simply the fact that in the latter the
affine connection $\nabla$ is not supposed to be symmetric. As a consequence,
the affine connection $\nabla$ is no longer a Levi-Civita connection and for
this reason affine geodesics do not coincide in general with metrical geodesics.

Since we are primarily interested in the embedding problem in the context of
spaces \ with torsion, in the next section we shall briefly examine the theory
of submanifolds in Riemann-Cartan geometry. As we shall see, the basic
mathematical facts are\ still simple extensions of the Riemannian case.

\section{Submanifolds and isometric embeddings in spaces with torsion}

We need first to review some basic concepts of the theory of Riemannian submanifolds.

\textbf{Definition}. \textbf{ } Let $(M,g,\nabla)$ \ and $(\overline{M}%
$,$\overline{g},\overline{\nabla})$ be Riemann-Cartan
differentiable\ manifolds of dimensions $m$ and $n=m+k$, respectively. A
differentiable map $f:M\rightarrow\overline{M\text{ }}$ is called an
\textit{immersion }if\ the\ differential $f_{\ast}:T_{P}(M)$ $\rightarrow
T_{f(P)}\overline{M}$ is injective for any $P\in M$. The number $k$ is called
the \textit{codimension }of $f$. We say that the immersion $f:M\rightarrow
\overline{M\text{ }}$ is \textit{isometric} at a point $P\in M$ if
$g(U,V)=\overline{g}($ $f_{\ast}(U),$ $f_{\ast}(V))$ for every $U,V$ in the
tangent space $T_{P}(M)$. If, in addition, $f$ is a homeomorphism onto $f(M)$,
then we say that $f$ is an \textit{embedding.} \bigskip If $M\subset
\overline{M}$ and the inclusion $i:M\subset\overline{M}$ $\rightarrow
\overline{M}$ is an embedding, then $M$ is\ called a \textit{submanifold} of
$\overline{M}$.

Let $f:M\rightarrow\overline{M\text{ }}$ be an embedding. We may, therefore,
identify $M$ with its image under $f$, so that we can regard $M$ as a
submanifold embedded in $\overline{M}$, with $f$ \ actually being the
inclusion map. Thus, we shall identify each vector $V\in T_{P}(M)$ with
$f_{\ast}(V)\in T_{f(P)}(\overline{M})$ and consider $T_{P}(M)$ as a subspace
of $T_{f(P)}(\overline{M}).$ In the vector space $T_{P}(\overline{M})$ the
metric $\overline{g}$ allows one to make the decomposition $T_{P}(\overline
{M})=T_{P}(M)\oplus T_{P}(M)^{\bot},$ where $T_{P}(M)^{\bot}$ is the
orthogonal complement of $T_{P}(M)$ $\subset T_{P}(\overline{M})$. That is,
for any vector $\overline{V}\in T_{P}(\overline{M})$, with $P\in M$, we can
decompose $\overline{V}$ into $\overline{V}=V+V^{\bot},$ $V\in T_{P}(M)$,
\ $V^{\bot}\in T_{P}(M)^{\bot}$.

Let us denote the connection on $\overline{M}$ by $\overline{\nabla}.$We now
can prove the following proposition.

\textbf{Proposition}. \textit{If }$V$\textit{ and }$U$\textit{ are local
vector fields on }$M$\textit{, and }$\overline{V}$\textit{ and }$\overline{U}%
$\textit{ are local extensions of these fields to }$\overline{M}$\textit{,
then the connection }$\nabla_{V}U$\textit{ compatible with the induced metric
on }$M$\textit{\ will be given by }%
\begin{equation}
\nabla_{V}U=(\overline{\nabla}_{\overline{V}}\overline{U})^{\top}\label{c}%
\end{equation}
\textit{where }$(\overline{\nabla}_{\overline{V}}\overline{U})^{\top}$\textit{
is the tangential component of }$\overline{\nabla}_{\overline{V}}\overline{U}%
$\textit{. }

Proof. It is not difficult to verify that $\nabla_{V}U$ as defined by
(\ref{c}) satisfies (\ref{affine1}), (\ref{affine2}) and (\ref{affine3});
hence our definition makes sense. Now consider the equation that expresses the
compatibility requirement between $\overline{\nabla}$ and $\overline{g}$ :%
\begin{equation}
\overline{V}[\overline{g}(\overline{U},\overline{W})]=\overline{g}%
(\overline{\nabla}_{\overline{V}}\overline{U},\overline{W})+\overline
{g}(\overline{U},\overline{\nabla}_{\overline{V}}\overline{W})\label{W1}%
\end{equation}
where $\overline{V}$, $\overline{U}$ , $\overline{W}$ \ $\in$ $T(\overline
{M}).$ Now, suppose that $\overline{V}$, $\overline{U}$ , $\overline{W}$ are
local extensions of the the vector fields $V,U,W$ to $\overline{M}.$ Clearly,
at a point $P\in M$, we have
\begin{equation}
\overline{V}[\overline{g}(\overline{U},\overline{W})]=V[\overline{g}%
(\overline{U},\overline{W})]=V[g(U,W)]\label{a}%
\end{equation}
where we have taking into account that the inclusion of $M$ into
$\overline{M\text{ }}$ is isometric. On the other hand, evaluating the first
term of the right-hand side of (\ref{W1}) at $\ P$ yields%
\begin{equation}
\overline{g}(\overline{\nabla}_{\overline{V}}\overline{U},\overline
{W})=\overline{g}((\overline{\nabla}_{\overline{V}}\overline{U})^{\top
}+(\overline{\nabla}_{\overline{V}}\overline{U})^{\bot},\overline
{W})=\overline{g}((\overline{\nabla}_{\overline{V}}\overline{U})^{\top
},\overline{W})=g((\overline{\nabla}_{\overline{V}}\overline{U})^{\top
},W)\label{b}%
\end{equation}
with an analogous expression for $\overline{g}(\overline{U},\overline{\nabla
}_{\overline{V}}\overline{W})$. From the above equations\ we finally obtain
\[
V[g(U,W)]=g((\overline{\nabla}_{\overline{V}}\overline{U})^{\top
},W)+g(U,(\overline{\nabla}_{\overline{V}}\overline{W})^{\top})
\]
From the Levi-Civita theorem extended to Riemann-Cartan manifolds, which
asserts the uniqueness of affine connection $\nabla$ in a Riemann-Cartan
manifold we conclude that \ the tangential component of $\overline{\nabla
}_{\overline{V}}\overline{U}$, \ evaluated at points of $M$, is, in fact, the
connection of $M$ compatible with the induced metric $g$ of $M.$

Since the embedding $f:M\rightarrow\overline{M\text{ }}$ induces a connection
$\nabla$ in $M$, given by (\ref{c}), we now turn our attention to the torsion
$T$ in $M$ defined by this induced connection. If $U$\textit{ }and \textit{
}$V$\textit{ are local vector fields on }$M$, then from (\ref{torsionless}%
)\ \ and (\ref{c})\ we get
\begin{equation}
T(U,V)=\nabla_{U}V-\nabla_{V}U-[U,V]=(\overline{\nabla}_{\overline{U}%
}\overline{V})^{\top}-(\overline{\nabla}_{\overline{V}}\overline{U})^{\top
}-[\overline{U},\overline{V}]^{\top} \label{inducedtorsion}%
\end{equation}
where $\overline{U}$ , $\overline{V}$ are local extensions of the the vector
fields $U,V$ to $\overline{M}$, and we have used the fact that at any point of
$M$ one has $[U,V]=[\overline{U},\overline{V}]$\ $=[\overline{U}^{\top
},\overline{V}^{\top}]\ =[\overline{U},\overline{V}]^{\top}$. Thus at any
point of $M$ the equation (\ref{inducedtorsion})\ becomes
\[
T(U,V)=(\overline{\nabla}_{\overline{U}}\overline{V}-\overline{\nabla
}_{\overline{V}}\overline{U}-[\overline{U},\overline{V}])^{\top}=\overline
{T}(\overline{U},\overline{V})^{\top}%
\]
We conclude therefore that the induced torsion $T$ on $M$ is nothing less than
the tangential component of the torsion $\overline{T}$ defined in
$\overline{M}$.

At this point consider the following question: Is it possible to have a purely
Riemannian submanifold $M$ isometrically embedded in a non-Riemannian space
$\overline{M}$ with a non-vanishing torsion? The answer is \ clearly
affirmative. Indeed, a submanifold $M$ embedded in space $\overline{M}$ with
torsion $\overline{T}$\ will be purely Riemannian if and only if the torsion
$T$ induced from $\overline{T}$ vanishes throughout $M$. From the above we see
that the necessary and sufficient condition for that is $\overline
{T}(\overline{U},\overline{V})^{\top}=0$, i.e., that the tangential component
of $\overline{T}$ of $\overline{M}$ vanishes identically.

To get further insight into the ideas developed above let us consider the case
in which the manifold $\overline{M}$ is foliated by a family of submanifolds
defined by $k$ equations $y^{A}=$constant \footnote{From now on lower case
Latin indices take value in the range $(0,1,...,(n+3))$, while Greek indices
run over $(0,1,2,3).$ The coordinates of a generic point $P$ of the manifold
$\overline{M}$ will be denoted by $y^{a}=(x^{\alpha},y^{4},...y^{n+3})$, where
$x^{\alpha}$ denotes the four-dimensional spacetime coordinates and
$y^{A}(A>3)$ refers to the $n$ extra coordinates of $P$.}, with the spacetime
$M$ corresponding to one of these manifolds $y^{A}=y_{o}^{A}=$constant. In
local coordinates $\left\{  y^{a}\right\}  $ of $\overline{M}$\ adapted to the
embedding it is not difficult to verify that the condition $T(U,V)=\overline
{T}(\overline{U},\overline{V})^{\top}=0$ implies $\overline{\mathcal{T}%
}_{.\alpha\beta}^{\lambda}=0$, where $\alpha,\beta,\lambda$ are tensorial
indices with respect to $M$. Therefore, if the components $\overline
{\mathcal{T}}_{.bc}^{\alpha}$ of the torsion tensor $\mathcal{T}$ \ vanishes
on $M$, then the geometry of the submanifold $M$ embedded in the
non-Riemannian bulk $\overline{M}$ is Riemannian.

It should be noted that the Riemannian character of spacetime $M$ embedded in
a Riemann-Cartan bulk $\overline{M}$ does not prevent the former from being
indirectly affected by the torsion of $\overline{M}$ . A nice illustration of
this point is given by the behaviour of geodesics near $M$. In \ the next
section we shall examine how a torsion field may affect the geodesic motion in
the case of a bulk with a warped product geometry. We shall be interested
particularly in the problem of classical confinement and stability of the
motion of particles and photons near the spacetime submanifold. \cite{Seahra,
Dahia}

\section{Geodesic motion in a Riemannian warped product space}

In this section let us consider the case where the geometry of the bulk
contains two special ingredients: a) It is a Riemannian manifold and \ b) its
metric has the structure of a warped product space \cite{Frolov}. As is well
known, the importance of warped product geometry is closely related to the
so-called braneworld scenario \cite{Roy}. Let us start with the investigation
of geodesics in warped product spaces, firstly considering the Riemannian case.

We define a warped product space in the following way. Let $(M,g)$ and $(N,h)$
be two\ Riemannian manifolds of dimension $m$ and $r$, with metrics $g$ and
$h,$ respectively. Suppose we are given a smooth function $f:N\rightarrow R$
(which will be\textbf{ }called the \textit{warping function}). \ We construct
a new Riemannian manifold by setting $\overline{M}=M\times N$ and endow
$\overline{M}$ with the metric $\overline{g}=e^{2f}g\oplus k$. We a view to
future application, we shall take $M=M^{4}\mathbb{\ }$and $N=R$, where $M^{4}$
denotes a four-dimensional (4D) Lorentzian manifold with signature $(+---)$
(henceforth referred to as \textit{spacetime}). In local coordinates
$\{y^{a}=(x^{\alpha},y^{4})\}$ the line element corresponding to the metric
$\overline{g}$ will be written as \footnote{Throughout this section Latin
indices take values in the range (0,1,...4) while Greek indices run from
(0,1,2,3).}
\[
dS^{2}=\overline{g}_{ab}dy^{a}dy^{b}%
\]

The equations of geodesics in the five-dimensional (5D) space $\overline{M}$
will be given by%
\begin{equation}
\frac{d^{2}y^{a}}{d\lambda^{2}}+^{(5)}\Gamma_{bc}^{a}\frac{dy^{b}}{d\lambda
}\frac{dy^{c}}{d\lambda}=0, \label{geodesics5D}%
\end{equation}
where $\lambda$ is an affine parameter and $^{(5)}\Gamma_{bc}^{a}$ denotes the
5D Christoffel symbols $^{(5)}\Gamma_{bc}^{a}=\frac{1}{2}\overline{g}%
^{ad}\left(  \overline{g}_{db,c}+\overline{g}_{dc,b}-\overline{g}%
_{bc,d}\right)  $. Denoting the fifth coordinate $y^{4}$ by $y$ and the
remaining coordinates $y^{\mu}$\ (the spacetime\textit{\ }coordinates) by
$x^{\mu}$, i.e. $y^{a}=(x^{\mu},y)$, we can easily show that the "4D part" of
the geodesic equations (\ref{geodesics5D}) can be rewritten in the form
\begin{equation}
\frac{d^{2}x^{\mu}}{d\lambda^{2}}+^{(4)}\Gamma_{\alpha\beta}^{\mu}%
\frac{dx^{\alpha}}{d\lambda}\frac{dx^{\beta}}{d\lambda}=\xi^{\mu},
\label{4Dpart}%
\end{equation}
where
\begin{align}
\xi^{\mu}  &  =-^{(5)}\Gamma_{44}^{\mu}\left(  \frac{dy}{d\lambda}\right)
^{2}-2^{(5)}\Gamma_{\alpha4}^{\mu}\frac{dx^{\alpha}}{d\lambda}\frac
{dy}{d\lambda}\nonumber\\
&  -\frac{1}{2}\overline{g}^{\mu4}\left(  \overline{g}_{4\alpha,\beta
}+\overline{g}_{4\beta,\alpha}-\overline{g}_{\alpha\beta,4}\right)
\frac{dx^{\alpha}}{d\lambda}\frac{dx^{\beta}}{d\lambda},
\end{align}
and $^{(4)}\Gamma_{\alpha\beta}^{\mu}=\frac{1}{2}\overline{g}^{\mu\nu}\left(
\overline{g}_{\nu\alpha,\beta}+\overline{g}_{\nu\beta,\alpha}-\overline
{g}_{\alpha\beta,\nu}\right)  $.

Let us now turn our attention to the five-dimensional brane-world scenario,
where the bulk corresponds to the five-dimensional manifold $\overline{M}$,
which, as in the previous section, is assumed to be foliated by a family of
submanifolds (in this case, hypersurfaces) defined by the equation $y=$ constant.

It turns out that the geometry of a generic hypersurface $\Sigma$, say
\ $y=y_{0},$ will be determined by the induced metric $g_{\alpha\beta
}(x)=\overline{g}_{\alpha\beta}(x,y_{0})$. Thus, on the hypersurface we have%
\[
ds^{2}=\overline{g}_{\alpha\beta}(x,y_{0})dx^{\alpha}dx^{\beta}.
\]
We see then that the quantities $^{(4)}\Gamma_{\alpha\beta}^{\mu}$ which
appear on the left-hand side of Eq. (\ref{4Dpart}) are to be identified with
the Christoffel symbols associated with the induced metric in the leaves of
the foliation defined above.

Let us restrict ourselves to the class of warped geometries given by the
following line element
\begin{equation}
dS^{2}=e^{2f}g_{\alpha\beta}dx^{\alpha}dx^{\beta}-dy^{2}, \label{warped}%
\end{equation}
where $f=f(y)$ and $g_{\alpha\beta}=$ $g_{\alpha\beta}(x)$. For this metric it
is easy to see\footnote{In the above calculation we have used the fact that
the matrix $g_{\alpha\beta}$ has an inverse $g^{\alpha\beta}$, that is,
$\ g^{\mu\beta}g_{\beta\nu}=\delta_{\nu}^{\mu}$. This may be easily seen since
by definition $\det g=-\det\overline{g}\neq0$.} that $^{(5)}\Gamma_{44}^{\mu
}=0$ and $^{(5)}\Gamma_{4\nu}^{\mu}=\frac{1}{2}\overline{g}^{\mu\beta
}\overline{g}_{\beta\nu,4}=f^{\prime}\delta_{\nu}^{\mu}$, where prime denotes
derivative with respect to $y$. Thus in the case of the warped product space
(\ref{warped}) the right-hand side of Eq. (\ref{4Dpart}) reduces to $\xi^{\mu
}=-2f^{\prime}\frac{dx^{\mu}}{d\lambda}\frac{dy}{d\lambda}$ and the 4D part of
the geodesic equations becomes
\begin{equation}
\frac{d^{2}x^{\mu}}{d\lambda^{2}}+^{(4)}\Gamma_{\alpha\beta}^{\mu}%
\frac{dx^{\alpha}}{d\lambda}\frac{dx^{\beta}}{d\lambda}=-2f^{\prime}%
\frac{dx^{\mu}}{d\lambda}\frac{dy}{d\lambda}. \label{4Dwarped}%
\end{equation}
On the other hand the geodesic equation for the fifth coordinate $y$ in the
warped product space becomes%

\begin{equation}
\frac{d^{2}y}{d\lambda^{2}}+f^{\prime}e^{2f}g_{\alpha\beta}\frac{dx^{\alpha}%
}{d\lambda}\frac{dx^{\beta}}{d\lambda}=0. \label{lwarped}%
\end{equation}
If the 5D geodesics are assumed to be timelike $\left(  \overline{g}_{ab}%
\frac{dy^{a}}{d\lambda}\frac{dy^{b}}{d\lambda}=1\right)  $, then we can
decouple the above equation from the 4D spacetime coordinates to obtain
\begin{equation}
\frac{d^{2}y}{d\lambda^{2}}+f^{\prime}\left(  1+\left(  \frac{dy}{d\lambda
}\right)  ^{2}\right)  =0. \label{5Dmotion}%
\end{equation}
Similarly, to study the motion of photons in 5D, we must consider the null
geodesics $\left(  \overline{g}_{ab}\frac{dy^{a}}{d\lambda}\frac{dy^{b}%
}{d\lambda}=0\right)  $, in which case Eq. (\ref{lwarped}) becomes
\begin{equation}
\frac{d^{2}y}{d\lambda^{2}}+f^{\prime}\left(  \frac{dy}{d\lambda}\right)
^{2}=0. \label{5Dphoton}%
\end{equation}

\bigskip Equations (\ref{5Dmotion}) and (\ref{5Dphoton}) are ordinary
differential equations of second-order which, in principle, can be solved if
the function $f^{\prime}=f^{\prime}(y)$ is known. A qualitative picture of the
motion in the fifth dimension may be obtained without the need to solve
(\ref{5Dmotion}) and (\ref{5Dphoton}) analytically \cite{Dahia}. This is done
by defining the variable $q=\frac{dy}{d\lambda}$ and then investigating the
autonomous dynamical system \cite{Andronov}
\begin{align}
\frac{dy}{d\lambda}  &  =q\label{dynamical}\\
\frac{dq}{d\lambda}  &  =F(q,y)
\end{align}
with $F(q,y)=-f^{\prime}(\epsilon+q^{2})$, where$\ \epsilon=1$ in the case of
(\ref{5Dmotion}) (corresponding to the motion of particles with nonzero rest
mass) and $\epsilon=0$ in the case of (\ref{5Dphoton}) (corresponding to the
motion of photons). In the investigation of dynamical systems a crucial role
is played by their \textit{equilibrium points}, which in the case of system
(\ref{dynamical}) are given by $\frac{dy}{d\lambda}=0$ and $\frac{dq}%
{d\lambda}=0$. These solutions, corresponding to fixed points in the phase
space of the system, represent curves that lie entirely in a a certain
hypersurface $\Sigma$ of the foliation previoulsy mentioned.\ The knowledge of
these points together with their stability properties provides a great deal of
information on the types of behaviour allowed by the system. An detailed
investigation of the qualitative behaviour of the solutions to the above
system was carried out in the cases when the five-dimensional $\overline{M}$
is Riemannian \cite{Dahia}, and when $\overline{M}$ is Weylian \cite{Gomez}.
In the next section we shall turn our attention to the case when $\overline
{M}$ $\ $is a Riemann-Cartan manifold, i.e., when $\overline{M}$ has torsion.

One of the motivations for studying the geodesic motion in the presence of
torsion is the following. As is well known, in the brane-world scenario the
stability of the confinement of matter fields at the quantum level is made
possible by assuming an interaction of matter with a scalar field. An example
of how this mechanism works is nicely illustrated \ by a field-theoretical
model devised by Rubakov, in which fermions may be trapped to a brane by
interacting with a scalar field that depends only on the extra dimension
\cite{Rubakov}. On the other hand, the kind of confinement we are concerned
with is purely geometrical, and that means the only force acting on the
particles is the gravitational force. In a purely classical (non-quantum)
picture, one would like to have effective mechanisms, other than the presence
of a quantum scalar field, to constrain massive particles to move on
hypersurfaces in a stable way. Two possibilities of implement such a program
have already been studied. One is to assume a direct interaction between the
particles and a physical scalar field \cite{Dahia1}. Following this approach
it has been shown that stable confinement in a thick brane is possible by
means of a direct interaction of the particles with a scalar field through a
modification of the Lagrangian of the particle. A second approach would appeal
to pure geometry: for instance, modelling the bulk with a Weyl geometrical
structure. In this case it is the Weyl field that provides the mechanism
necessary for confinement and stabilization of the motion of particles in the
brane \cite{Gomez}. At this stage, we would like to know whether classical
confinement of particles and photons could also be obtained by using a torsion
field, that is, by allowing for the bulk to have a Riemann-Cartan geometry.
This is the question we shall deal with in the next section.

\section{Geodesic motion in the presence of torsion}

When the five-dimensional embedding space $\overline{M}=M^{5}$ is a warped
product space endowed with a torsion field it is not difficult to verify, by
putting $^{(5)}\Gamma_{bc}^{a}=\left\{  _{bc}^{a}\right\}  -K_{\;bc}^{a}$ into
(\ref{geodesics5D}) and noting that $K_{\;a4}^{4}=0$, that the motion of a
massive particle in the fifth dimension is given by the equation%
\begin{equation}
\frac{d^{2}y}{d\lambda^{2}}+f^{\prime}\left(  1+\left(  \frac{dy}{d\lambda
}\right)  ^{2}\right)  =K_{\;(\alpha\beta)}^{4}\frac{dx^{\alpha}}{d\lambda
}\frac{dx^{\beta}}{d\lambda}+K_{\;4\alpha}^{4}\frac{dx^{\alpha}}{d\lambda
}\frac{dy}{d\lambda},\label{torsion1}%
\end{equation}
where $K_{\;(\alpha\beta)}^{4}$ denotes the symmetric part of $K_{\;\alpha
\beta}^{4}$. This equation is the equivalent of (\ref{5Dmotion}) for a
five-dimensional bulk is a Riemann-Cartan manifold, i.e., when $M^{5}$ has a
non-vanishing torsion. In this form Eq. (\ref{torsion1}) does not allow us, in
general, to study the motion of the particle along the extra dimension
decoupled from its motion in the four-dimensional spacetime. Nevertheless,
there are some particular cases in which the five-dimensional motion is
independent of the remaining dimensions. One case is, of course, when both
$K_{\;(\alpha\beta)}^{4}$ and\ $K_{\;\alpha4}^{4}=\overline{\mathcal{T}%
}_{\;4\alpha}^{4}$ vanish. In this situation the torsion field does not
influence the motion of the particle along the fifth dimension. A more
interesting case, however, is when the symmetric part of the components
$K_{\;(\alpha\beta)}^{4}$ of the contorsion tensor is chosen proportional to
$h_{\alpha\beta}=$ $e^{2f}g_{\alpha\beta}$, the induced metric on $M$. For
instance, if we set $K_{\;(\alpha\beta)}^{4}=\psi(y)e^{2f}g_{\alpha\beta}$ and
$K_{\;4a}^{4}=0$, then the equation (\ref{torsion1}) becomes
\[
\frac{d^{2}y}{d\lambda^{2}}+\left(  1+\left(  \frac{dy}{d\lambda}\right)
^{2}\right)  \left(  f^{\prime}(y)-\psi(y)\right)  =0
\]
The presence of the function $\psi(y)$ modifies the dynamical system
(\ref{dynamical}) leading to a new picture of the phase plane, where the
equilibrium points and the stability properties of the solutions may
completely change. To give an illustration, let us consider that the
five-dimensional Riemannian space $\overline{M}$ is endowed with a
Mashhoon-Wesson-type metric \cite{Mashhoon}%
\begin{equation}
dS^{2}=\frac{\Lambda^{2}}{3}y^{2}g_{\alpha\beta}dx^{\alpha}dx^{\beta}%
-dy^{2}.\label{Mashhoon}%
\end{equation}
It has been shown \cite{Gomez} that in this case \ there is no confinement of
particles in the hypersurfaces $y=const$. Now let us introduce a torsion field
such that it gives rise to a \ contorsion field\ $K_{\;\alpha\beta}^{\lambda}%
$\ given by
\begin{equation}
K_{\;\alpha\beta}^{4}=\psi(y)g_{\alpha\beta}+L_{\alpha\beta}%
\label{contorsion2}%
\end{equation}
with $L_{\;\;\alpha\beta}^{4}$ antisymmetric in the indices $\alpha,\beta$.
With this choice let us show that we can set up a confinement mechanism
induced by the contorsion only. With this objective in mind let us set
$\psi(y)=\frac{1}{y}-a(y-y_{0})$, where $a$ is a constant. It is an easy task
to show that with such choice all timelike geodesics of $M=M^{4}$ will remain
confined in $M^{4}$. Moreover, the stability of the confinement in this case
is entirely governed by the sign of the constant $a$. All these results can be
obtained from an analysis of the dynamical system (\ref{dynamical}) \ carried
out in the neighbourhood of the equilibrium points,\ now with the function
$F(q,y)$ modified due to the presence of torsion. The new dynamical system to
be studied is thus
\begin{align}
\frac{dy}{d\lambda} &  =q\label{dynamical2}\\
\frac{dq}{d\lambda} &  =-(1+q^{2})(f^{\prime}(y)-\psi(y))
\end{align}
For the Mashhoon-Wesson metric $f^{\prime}(y)=\frac{1}{y}$, so if the torsion
field is not present, then it is clear that there are no equilibrium points
(because $f^{\prime}(y)$ has no roots). That means no confinement of massive
particles is possible at the hypersurfaces $y=y_{0}=const$. On the other hand,
if the torsion field is \ "turned on", then (\ref{dynamical2}) becomes
\begin{figure}[t]
\centering
\includegraphics[width=.5\textwidth,angle=-90]{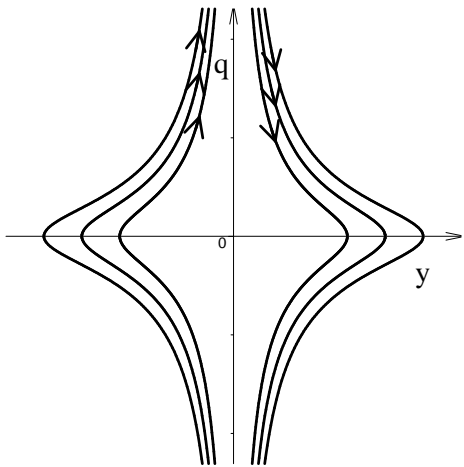}\caption{{\footnotesize In
the absence of torsion there are no equilibrium points. The particles go
through the leaves without being trapped by the hypersurface $\Sigma$.}}%
\label{figura1}%
\end{figure}

\begin{figure}[t]
\centering
\includegraphics[width=.5\textwidth,angle=-90]{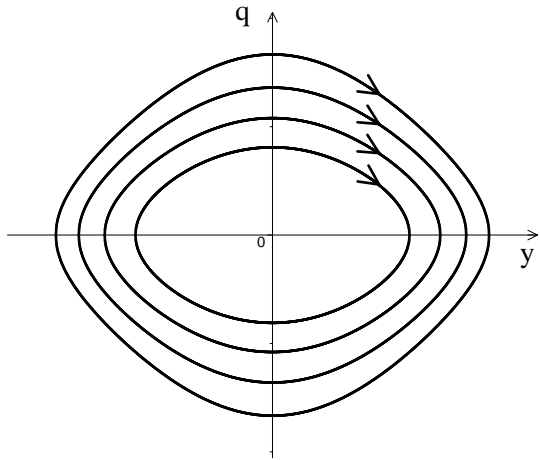}\caption{{\footnotesize The
effect of the torsion field is to "close" the paths in phase plane, with the
appearance of a trapping hypersurface, pictured in the diagram as the
equilibrium point $E$.}}%
\label{figura1}%
\end{figure}%

\begin{align}
\frac{dy}{d\lambda}  &  =q\label{dynamical3}\\
\frac{dq}{d\lambda}  &  =a(y-y_{0})(1+q^{2})
\end{align}
In regard to the dynamical system above we now have an equilibrium point $E$
at $q=0,y=y_{0}$. This is a solution that corresponds to the worldline of a
massive particle trapped under the action of the torsion field at the
hypersurface $y=y_{0}$. \ It is straightforward to verify that if $a>0$, then
the equilibrium point $E$ corresponds to a \textit{center}. In other words,
the solutions near $E$ have the topology of a circle in the phase portrait of
the dynamical system (\ref{dynamical3}). The closed curves thus describe the
motion of particles oscillating about the hypersurface $y=y_{0}$.\ In this
case we are in the presence of a kind of confinement where particles lying
near the $y=y_{0}$\ will oscillate about it, entering and leaving the
hypersurface indefinitely. On the other hand, if $a<0$, $E$ is a\textit{
saddle point}. In this case, although the particle is still constrained\ by
the torsion field to move on $\Sigma$, this sort of confinement is highly
unstable: Almost any small perturbation on the \ fifth-dimensional motion of
the particle will cause it to be expelled from $\Sigma$.

Before concluding this section, let us consider the question whether
(\ref{contorsion2})\ represents a possible choice, i.e., a legitimate choice
for the components of contorsion field. In order to answer this question just
choose any torsion tensor having the following components:%

\begin{equation}
\mathcal{T}_{\;\beta4}^{\alpha}=-\delta_{\beta}^{\alpha}\psi(y),\mathcal{T}%
_{\;\alpha\beta}^{4}=2L_{\alpha\beta}\text{.}\label{example}%
\end{equation}
A simple calculation is sufficient to convince ourselves that (\ref{example})
will lead to\ (\ref{contorsion2}).

\section{Final Remarks}

In the recent years there has been a renewed interest in a certain class of
higher-dimensional spacetime theories which start from the following
assumptions: a) our spacetime is viewed as four-dimensional \ Riemannian
hypersurface embedded in a five-dimensional Riemannian\ manifold (the bulk);
b) the geometry of the higher-dimensional space is characterized by a warped
product space; c) fermionic matter is confined to the hypersurface by means of
an interaction of the fermions with a scalar field which depends only on the
extra dimension. In this scenario we have considered the possibility of
describing the five-dimensional space by a non-Riemannian geometry, namely a
Riemann-Cartan geometry, in which a new degree of freedom, the torsion,
appears. We have shown that for a class of torsion fields, the geometry
induced on four-dimensional spacetime has a Riemannian structure. This means
that it is possible to embed isometrically\ a Riemannian spacetime into a
Riemann-Cartan five-dimensional bulk with non-vanishing torsion. We also have
shown that confinement and stability properties of geodesics near the brane
may be affected by the torsion. As an illustration of this fact, we have
considered the case of a five-dimensional warped product space and have
constructed\ a classical analogue of the quantum confinement by considering a
very special case of torsion field. In a certain sense, this purely
geometrical field, which has a purely geometrical nature, is able to replace
the quantum scalar field that is usually responsible for the confinement in
field-theoretical models \cite{Rubakov}.\ 

Another comment concerning the embedding of the spacetime in spaces with
torsion is the following. In the induced-matter approach an energy-matter
tensor describing macroscopic matter is generated geometrically from the
Einstein field equations in vacuum. \ It is now well understood that the
question whether any energy-momentum tensor\ $T_{\alpha\beta}$\ can be
generated in this way is equivalent to know whether any solution of the
Einstein equations for\ a prescribed $T_{\alpha\beta}$ can be locally embedded
in some five-dimensional Ricci-flat space. As it happens, the answer to this
question is given by the Campell-Magaard theorem, which states that any
$n$-dimensional Riemannian manifold can be locally and isometrically embedded
in a $n+1)$-dimensional Ricci-flat Riemannian space \cite{campbell}.
Transposing these ideas to the Einstein-Cartan theory of gravity, one would
naturally wonder whether spin could also be generated, or induced, in the same
manner, from a higher-dimensional space. The results obtained in Sec. 3, allow
us to draw some conclusions in this respect. One is that if the bulk
$\overline{M}$\ is a torsionless space ( hence not sourced by matter with
spin), then it is not possible to generate spin geometrically (through
dimensional reduction) in the four-dimensional spacetime $\overline{M}$. A
second conclusion is that, in general, spin in four dimensions may be
generated from five dimensions, but in some particular cases the bulk does not
transfer spin to four-dimensions.

In a certain sense, the present article is a follow up of previous work, where
basically the same questions treated here were examined in the context of
another non-Riemannian setting, namely that of the geometry of Weyl
\cite{Gomez}. In fact, we may regard these works as part of more general
program of research whose underlying idea is to highlight and explore the role
non-Riemannian geometries may play in the development of novel frameworks for
modern higher-dimensional spacetime theories.

\section{\bigskip Acknowledgement}

The authors would like to thank CNPq-FAPESQ (PRONEX) and FAPES for financial
support. L. F. P. da Silva acknowledges FAPEAM for \ a grant.

\end{document}